# Mechanical Vibrational Relaxation of NO Scattering from Metal and Insulator Surfaces: When and Why Are They Different?

Rongrong Yin and Bin Jiang[*]

*Hefei National Laboratory fo1r Physical Science at the Microscale, Department of Chemical Physics, Key Laboratory of Surface and Interface Chemistry and Energy Catalysis of Anhui Higher Education Institutes, University of Science and Technology of China, Hefei, Anhui 230026, China*

NO scattering from metallic and insulating surfaces represent contrasting benchmark systems for understanding energy transfer at gas-surface interface. Strikingly different behaviors of highly vibrationally excited NO scattered from Au(111) and LiF(001) were observed and attributed to disparate electronic structures between metals and insulators. Here, we reveal an alternative mechanical origin of this discrepancy by comparative molecular dynamics simulations with globally accurate adiabatic neural network potentials of both systems. We find that highly-vibrating NO can reach for the high dissociation barrier on Au(111), by which vibrational energy can largely transfer to translation/rotation and further dissipate into substrate phonons. This mechanical energy transfer channel is forbidden in the purely repulsive NO/LiF(001) system or for low-vibrating NO on Au(111), where molecular vibration is barely coupled to other degrees of freedom. Our results emphasize that the initial state and potential energy landscape concurrently influence the mechanical energy transfer dynamics of gas-surface scattering.

Energy exchange among molecular and surface degrees of freedom (DOFs) contributes fundamentally to various physical and chemical phenomena at gas-solid interfaces[1]. Vibrational energy transfer is of particular importance as vibrational coordinates are intrinsically relevant to forming and breaking of bonds in surface chemical reactions. This has motivated a large body of state-to-state molecular scattering experiments at solid surfaces demonstrating the most detailed dynamics on how molecular vibration couples with other DOFs[2-7]. Such state-resolved data have guided the development of theoretical models towards a predictive understanding of molecule-surface interactions[8-13].

NO scattering from solid surfaces is one of the best studied examples concerning vibrational energy transfer at gas-surface interfaces[14-25], due mainly to the seminal contributions of Wodtke and coworkers[26,27]. One of the most striking observations was the multi-quantum vibrational relaxation of highly vibrationally excited NO($v_i$=15) scattered from Au(111)[14], compared to the large vibrational elasticity of NO($v_i$=12) scattered from LiF(001)[23]. This discrepancy was attributed to the contrasting band structures of metals and insulators[14], as the mechanical (electronically adiabatic) gas-surface vibrational relaxation of small molecules was commonly recognized to be inefficient given the mismatch of molecular vibration and surface phonon frequencies[22,28]. That is to say, metals are favorable for electron transfer to NO and efficient vibration-to-electron coupling, while insulators would effectively turn off this electronically non-adiabatic vibrational relaxation channel. This concept was later supported by several different theoretical models[29-33]. In particular, the multistate-based independent electron surface hopping (IESH) model developed by Tully and coworkers[30] qualitatively reproduced the multi-quantum vibrational relaxation of NO($v_i$=15) on Au(111) at a low incidence energy ($E_i$=0.05 eV). In comparison, adiabatic dynamics calculations on the same ground state potential energy surface (PES) resulted in large vibrational elasticity[31] as observed in previous experimental[22,23,34] and theoretical[28] findings for NO on LiF(001), implying similarly inefficient mechanical vibrational energy transfer in both systems.

Nevertheless, the IESH model failed to capture some more recently discovered features in the NO/Au(111) system, such as the incidence energy[19,21] and orientation[25] dependence of vibrational relaxation and the final translational energy distributions[21]. These failures were attributed to the inaccuracy of the empirical function parameterized adiabatic PES used in IESH simulations[19], which have been largely remedied by a more accurate machine learned adiabatic NO/Au(111) PES developed by us based on thousands of density functional theory (DFT) data points[35]. Combining this PES with the orbital-dependent electronic friction tensor accounting for non-adiabatic effects[36], diverse experimental observations have been reasonably reproduced. This new PES predicted a surprisingly large amount of vibrational energy of the highly-vibrating NO dissipated mechanically into the gold surface[35], suggesting that the efficiency of mechanical gas-surface vibrational relaxation is not always low and is very dependent on the adiabatic PES and the underlying electronic structure description. In this Letter, we develop a new high-quality machine learned PES for NO scattering from LiF(001) fit to numerous DFT data. With the state-of-the-art adiabatic PESs for both systems, we are able to elucidate how the observed distinct vibrational inelasticity of NO scattered from metals and insulators is related to their adiabatic PESs and under what conditions the molecular vibrational energy is able to flow into substrate phonons despite the mismatch of their frequencies.

To construct the PES, periodic DFT calculations were performed using the Vienna ab Initio Simulation Package (VASP)[37,38] with the PW91 functional[39]. The LiF(001) surface was modeled with a four-layer slab in a 2×2 supercell, in which the top two layers were movable. The plane-wave basis set was truncated at 400 eV and Brillouin-zone was sampled by a 5×5×1 Monkhorst-Pack $k$-points mesh. To map out the PES including both molecular and surface DOFs, we collected 3608 data points using an adaptive sampling strategy[40] that fully cover the configuration space relevant to highly vibrationally excited NO scattering from LiF(001). Both energies and forces were trained using our recently proposed embedded atom neural network (EANN) approach[41-43]. The NO/Au(111) PES used here was also retrained with EANN using the DFT data in Ref. [35]. To calculate the state-to-state scattering probabilities of NO and account for the large energy exchange between the heavy molecule and the substrate, which is intractable at present by a fully-coupled quantum dynamics method, we carried out quasi-classical trajectory (QCT) calculations with the initial and final conditions of NO molecules quantized semiclassically. Such a treatment has shown to be reliable for diatomic scattering from surfaces, e.g. for molecular hydrogen[9]. More details about the EANN PES and QCT simulations are given in the Supplemental Material (SM).[44]

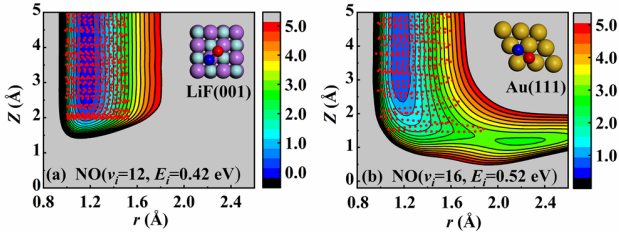

FIG. 1. Two-dimensional cuts of (a) the NO/LiF(001) and (b) the NO/Au(111) PESs as a function of the N-O distance ($r$) and the molecular height ($Z$) above the surface, with other coordinates (defined in Fig. S1) fixed at the adsorption state on LiF(001) and the dissociation transition state on Au(111). A representative trajectory (red dots) of each system is projected on top of the corresponding PES.

Let us first look at the first-principles-determined potential energy landscapes of the NO/Au(111) and NO/LiF(001) systems. As displayed in Fig. 1 and Fig. S1, the biggest difference between the two PESs is that NO is dissociative on Au(111) with a high barrier of ~2.86 eV with its bond length elongated to ~1.89 Å, while is absolutely non-dissociative and repulsive on LiF(001). Additionally, the NO molecular adsorption energy is higher on Au(111) than on LiF(001) (-0.39 eV versus -0.06 eV). Accordingly, the NO/Au(111) PES is much more anisotropic and corrugated than the NO/LiF(001) PES, as displayed in Fig. S2-4. These differences clearly reflect a stronger interaction of NO with a metal than with an ionic crystal.

Keeping this in mind, we discuss the adiabatic energy transfer dynamics in these two systems in similar conditions. Fig. 2 compares the calculated final vibrational state ($v_f$) distributions of highly vibrationally excited NO molecules scattered from both surfaces, with available experimental data[14,23] at surface temperature ($T_s$) of ~300 K. Although electronically non-adiabatic effects are not included, we see an obvious difference in the two cases. NO($v_i$=16, $E_i$=0.52 eV) scattering from Au(111) undergoes multi-quantum vibrational relaxation yielding a broad vibrational state distribution down to $v_f \approx 2$ and peaking at $v_f \approx 10$. This amounts roughly to half of the experimentally measured vibrational relaxation[21]. Our recent work showed that the agreement with experiment in this and other conditions can be greatly improved further if hot-electron effects were taken into account by the electronic friction theory with the orbital-dependent friction tensor and reasonable corrections to the Markovian approximation[36]. On the contrary, our calculations predict absolute vibrational elasticity of NO($v_i$=12, $E_i$=0.42 eV) scattered from LiF(001), in good accord with the high survival probability observed in experiment. Apparently, the much more significant vibrational relaxation of highly vibrationally excited NO from Au(111) than from LiF(001) cannot be explained by the direct vibration-phonon coupling. Fig. S5 shows that phonon frequencies of LiF(001) are generally several times higher than those of Au(111), relatively closer to the NO vibrational frequency (1906 cm$^{-1}$). This should enable faster vibrational relaxation rate of NO on LiF(001) in terms of the energy gap law for vibrational energy transfer[28].

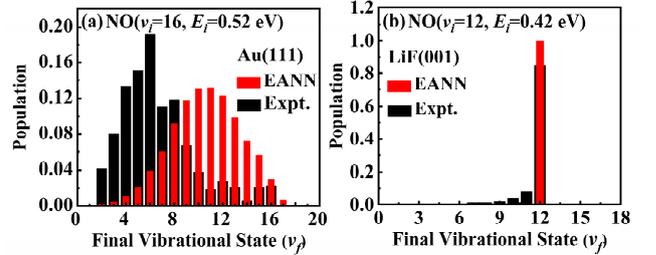

FIG. 2. Comparison of vibrational state distributions of highly vibrationally excited NO scattered from (a) Au(111) and from (b) LiF(001), including the adiabatic QCT results on the EANN PES (red) and experimental data (black). Initial conditions are selected mimicking the corresponding experiments.

Different adiabatic energy transfer dynamics in the two cases can be further seen in the energy partitioning of scattered molecules. Table I compares the mean values of the rotational ($<\Delta E_{rot}>$), vibrational ($<\Delta E_{vib}>$), translational ($<\Delta E_{trans}>$) and total energy loss ($<\Delta E_{tot}>$) of NO scattering from LiF(001) and Au(111), defined by the final energy minus the initial energy of NO. Although there is negligible vibrational energy loss of NO($v_i$=12) on LiF(001), about half of the translational energy is lost to the lattice, along with slight rotational excitation. This indicates that molecular vibration is barely coupled to other molecular and surface DOFs, originating from the purely repulsive, nearly isotropic and flat PES in this system (see Fig. 1 and S2-4). To illustrate this more explicitly, a representative trajectory of NO($v_i$=12, $E_i$=0.42 eV) scattering from LiF(001) is projected onto Fig. 1a. The vibrational motion is completely orthogonal to translation, whose amplitude is basically unperturbed upon the direct scattering process. Fig. 3a further shows that the mechanical energy exchange between molecular translation and the LiF lattice occurs upon a short contact with the surface. The NO molecule quickly converts all of its translational energy to potential energy at the repulsive wall, then gets only half of kinetic energy back when recoiling from the surface, resulting in a net translational energy loss to surface phonons. Interestingly, this calculated vibrational elasticity of NO scattering from LiF(001) is insensitive to the incidence energy, initial state, surface temperature, and even density functional (Fig. S6), as long as NO undergoes more or less the same repulsive force. Indeed, we find again no vibrational de-excitation of NO($v_i$=1) scattered from LiF(001) at $E_i$=0.31 eV, in line with the large survival probability in experiment (~0.9±0.1) in the same condition[22]. The average final translational energy of scattered molecules is ~0.21 eV, which compares well with the experimental value (~0.19 eV)[22]. The measured narrow angular distributions and rotational distributions[22,23] are also reasonably reproduced (Fig. S7). These results suggest that the scattering dynamics of NO from LiF(001) is well described by the new PES.

Table I. Average rotational, vibrational, translational, and total energy losses (in eV) of NO before and after scattering from LiF(001) and Au(111) in various initial conditions (see text).

| Mean energy loss (eV) | LiF(001) | | Au(111) | |
|---|---|---|---|---|
| | NO($v_i$=12) $E_i$=0.42 eV | NO($v_i$=1) $E_i$=0.31 eV | NO($v_i$=16) $E_i$=0.52 eV | NO($v_i$=1) $E_i$=0.31 eV |
| $<\Delta E_{rot}>$ | 0.035 | 0.037 | 0.40 | 0.055 |
| $<\Delta E_{vib}>$ | -0.0010 | -0.0013 | -1.0 | -0.0020 |
| $<\Delta E_{trans}>$ | -0.24 | -0.10 | -0.036 | -0.13 |
| $<\Delta E_{tot}>$ | -0.20 | -0.063 | -0.65 | -0.077 |

In sharp contrast, NO($v_i$=16, $E_i$=0.52 eV) scattered from Au(111) loses a substantial amount of vibrational energy ($<\Delta E_{vib}>$=-1.0 eV) and only a little translational energy ($<\Delta E_{trans}>$=-0.036 eV), flowing primarily into the lattice ($<\Delta E_{tot}>$=-0.65 eV) and partly to rotation ($<\Delta E_{rot}>$=0.40 eV). Unlike the NO/LiF(001) case, now the highly-vibrating NO molecule has an opportunity to lengthen and dissociate on Au(111)[35], during which the molecular vibration can gradually soften and thus couple with other DOFs. Fig. S8 demonstrates the mode softening from 1906 cm$^{-1}$ in the gas phase to 457 cm$^{-1}$ at the transition state along the minim energy path. Similar mode softening has effectively lowered the vibrationally adiabatic barrier height and enabled great vibrational enhancement of dissociative adsorption of polyatomic molecules at metal surfaces[55]. Its influence on vibrationally inelastic scattering is clearly seen via a representative trajectory of NO($v_i$=16→$v_f$=10) scattering from Au(111) in Fig. 1b and the corresponding kinetic energy evolution in each DOF as a function of time in Fig. 3d-e. In this case, before the first impact at the surface (Z>2.0 Å), the NO molecule has acquired additional translational energy by ~0.89 eV with its bond length slightly extended to ~1.65 Å. Note that the NO-Au(111) attraction may accelerate the NO molecule by at most ~0.39 eV (*i.e.*, the maximum adsorption well depth), but the excess translational energy gain must be transferred from vibration. As the molecule goes more deeply to the repulsive wall and elongates further to the transition state region, the vibration is more significantly softened and the translation is suppressed. Both the vibrational kinetic energy and translational energy largely decrease, accompanied with a rapid increase of the kinetic energy of surface atoms (of course also the potential energy of the system, which is however inseparable). Meanwhile, a large amount of energy is transferred to molecular rotation because the PES in this region features strong anisotropy and corrugation (Figs. S2-4) and the molecule would reorient to resemble the transition state geometry. Whereas the bond dissociation is unsuccessful and the molecule is eventually scattered back to the vacuum leaving the highly excited surface phonons and molecular rotation, yet much reduced vibrational excitation. Interestingly, even running calculations with all surface atoms fixed, we still see significant vibrational energy loss ($<\Delta E_{vib}>$=-0.71 eV) of NO scattered from Au(111) but with a net average energy gain in translation ($<\Delta E_{trans}>$=0.22 eV) and rotation ($<\Delta E_{rot}>$=0.49 eV). This signifies that a large fraction of translational energy and part of rotational energy transferred from vibration eventually flows to surface phonons in the strongly interacting region when the surface is relaxed (Table S1).

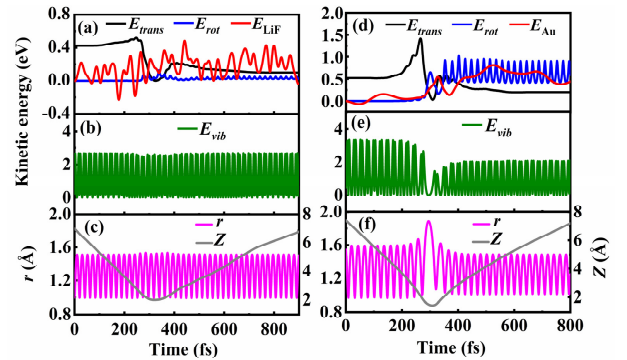

FIG. 3. Kinetic energy and geometric evolutions as a function of time during a representative scattering trajectory for NO($v_i$=12, $E_i$=0.42 eV) from LiF(001) (a-c) and NO($v_i$=16, $E_i$=0.52eV) from Au(111) (d-f), including the kinetic energy in NO translation ($E_{trans}$, black), rotation ($E_{rot}$, blue), and vibration ($E_{vib}$, green), the average kinetic energy of surface atoms relative to the initial value ($E_{LiF}$ and $E_{Au}$, red), the N-O distance (*r*, magenta), and the molecular height (*Z*, gray).

We emphasize that such efficient adiabatic vibrational energy transfer would happen for highly vibrationally excited states of NO on Au(111) only. Indeed, the scattering of NO($v_i$=1, $E_i$=0.31 eV) from Au(111) turns out to be more similar to that from LiF(001), giving rise to nearly no vibrational energy loss (see Table I). Moreover, the mean translational energy loss of NO is 0.13 eV on Au(111), comparable to that on LiF(001). As illustrated in Fig. S9, this is because that NO($v_i$=1) cannot reach the dissociation channel on Au(111) and directly scatters in the entrance channel, manifesting a similar behavior as on the repulsive LiF(001) surface. We have found that mechanical vibrational de-excitation becomes measurable for NO($v_i$=3) and increasingly significant as $E_i$ increase[35], as the molecule is more strongly affected by the force of the transition state region. The influence of the dissociation barrier on gas-surface scattering has been mainly discussed in systems with relatively low barriers and for low vibrational states[4,6,56,57]. Our results highlight that the scattering of highly vibrationally excited molecules from solid surfaces can be strongly affected by a dissociation channel even when the barrier is very high and customarily ignored. This explains the negligible vibrational relaxation in adiabatic dynamics calculations on the empirical PES without a dissociation barrier[31].

With this more quantitative understanding of vibrational relaxation at gas-surface interfaces, we turn to discuss the remaining discrepancy between theory and experiment for NO($v_i$=12) scattering from LiF(001). Defects in experimentally prepared LiF(001) samples have been invoked to explain the observed minor vibrational inelasticity of NO($v_i$=12)[23]. The influence of defects can be remarkable, for example, NO($v_i$=1) scattering was found completely vibrationally inelastic[58] from polished LiF(001) (with possibly high concentration of defects), while mostly vibrationally elastic[22] from well-cleaved LiF(001). By performing additional DFT calculations on defective LiF(001) surfaces with steps or ionic vacancies (Fig. S11-13), we find that those defects only moderately strengthen the NO-LiF binding but fail to support any dissociation transition states. Based on our analysis, however, the mechanical vibrational relaxation would be significant only if the repulsive PES was largely changed to enable some sort of chemical transformation. On the other hand, we find that a single Li$^+$ vacancy can significantly reduce the band gap of LiF(001) to ~2.40 eV, which may be accessible by NO($v_i$=12) with a vibrational energy of ~2.70 eV. Additional work is needed to learn whether this would cause some non-adiabatic vibrational energy loss of NO on imperfect LiF(001).

To summarize, we present a comparative study on the adiabatic energy transfer dynamics of NO scattering from Au(111) and LiF(001) revealed by high-dimensional global neural network PESs with DFT accuracy. We find almost exclusive vibrationally elastic products of NO($v_i$=12) scattered from LiF(001), in stark contrast with the multi-quantum relaxation NO($v_i$=16) from Au(111). However, the scattering of NO($v_i$=1) from both surfaces is vibrationally elastic and only translational energy leads to phonon excitation. This interesting discrepancy and similarity have been quantitatively rationalized by the potential energy landscape accessible by the impinging NO molecule. As long as the molecule reaches the dissociation barrier region, its vibration will be significantly softened and vibration-to-translation/rotation/phonon energy transfers are significant. While the low-vibrating NO molecule is rejected directly by the repulsive wall without experiencing the dissociative force and the vibrational softening on Au(111), rendering its similar behavior as on the purely repulsive LiF(001) surface. Our results not only reproduce well the experimental data for NO scattering on LiF(001), but also suggest a mechanical mechanism for the discrepant vibrational relaxation of highly-vibrating NO scattering from metallic and insulating surfaces. An apparent next step is to combine this mechanical and the electron-mediated non-adiabatic vibrational energy transfer channels, which could be both promoted by the same region of the PES[24], to fully understand the energy transfer dynamics of molecules like NO and CO at metal surfaces.

We appreciate the support from National Key R&D Program of China (2017YFA0303500), National Natural Science Foundation of China (22073089 and 22033007), Anhui Initiative in Quantum Information Technologies (AHY090200). We thank Prof. Alec Wodtke for some helpful discussion.

*bjiangch@ustc.edu.cn

# Supplemental Material

# Mechanical Vibrational Relaxation of NO Scattering from Metal and Insulator Surfaces: When and Why Are They Different?

Rongrong Yin and Bin Jiang[*]

*Hefei National Laboratory for Physical Science at the Microscale, Department of Chemical Physics, Key Laboratory of Surface and Interface Chemistry and Energy Catalysis of Anhui Higher Education Institutes, University of Science and Technology of China, Hefei, Anhui 230026, China*

[*]: corresponding author: bjiangch@ustc.edu.cn

# 1. Computational Details

## 1. 1 Density Functional Theory and Data Sampling

Total energies and atomic forces used to develop the high-dimensional reactive potential energy surfaces (PESs) in this work were obtained by density functional theory (DFT) within generalized gradient approximation (GGA), as implemented in Vienna Ab Initio Simulation Package (VASP)[37,38]. Since NO is an open-shell molecule, spin polarized calculations were performed. The LiF(001) surface was modeled by a four-layer slab in a 2 × 2 unit cell with the vacuum of 20 Å, where the top two layers were allowed to move. The widely-used PW91 functional[45] was chosen and the core-electron interactions were described with the projector augmented wave (PAW) method[46]. The plane-wave basis was truncated up to 400 eV and the Brillouin zone was sampled by a 5 × 5 × 1 Monkhorst-Pack $k$-points[47]. The convergence criterion was set as $10^{-6}$ eV for the electronic self-consistency iteration and 0.01 eV/Å for the ionic relaxation in geometric optimization. To study the influence of defects in NO scattering from LiF(001), we calculate the binding energies of NO on the pristine and several representative defective LiF(001) surfaces, $e.g.$ LiF(210), LiF(001) with a single $F^-$ vacancy, or a single Li+ vacancy. The LiF(001) surface with a single $F^-$ vancancy or a single $Li^+$ vancancy was modeled by remove a single $F^-$ atom or a single $Li^+$ atom from the first layer of LiF(001) slab. The LiF(210) was modeled by a nine-layer slab in a 2× 1 unit cell with the vacuum of 20 Å, where the top five layers were allowed to move. The DFT setups in these cases were similar as described above.

The NO + LiF(001) and NO + Au(111) coordinate systems for describing the state-to-state scattering process is schematically illustrated in Figure S1. In the Figure S1(a)(c), we have explicitly labelled N-O bond length ($r$), the distance of the molecular center of mass to surface ($Z$), the distance of N-surface ($Z_N$), the polar angle ($\theta$) and azimuthal angle ($\varphi$). To construct the high-dimensional potential energy surface (PES) for the NO/LiF(001) system, ab initio molecular dynamics (AIMD) simulations were initially performed to sample the configuration space. The clean surface configurations were taken from equilibrated AIMD snapshots of the clean surface at experimental surface temperature[23] ($T_s$=290 K and $T_s$=480 K), Non-equilibrium AIMD trajectories were initiated with NO being far above the LiF(001) surface ($Z$=6.0 Å). The initial lateral positions and orientations of the molecules were selected randomly covering the unit cell. The NO molecules with incident energies ($E_i$) of 0.30 eV and 0.75 eV were impinging at the surface along surface normal. These AIMD trajectories were propagated using the leapfrog algorithm in VASP with a time step of 0.5 fs, up to a maximum propagation time of 10 ps, if they were neither terminated as "reactive" whenever the N−O distance became larger than 2.5 Å, nor as "scattered" when the molecule was flying away from the surface by 6.0 Å.

We first selected ~500 points out of 5000 points from 20 non-equilibrium AIMD trajectories, according to their generalized Euclidian distances (GED) in terms of inter-nuclear distances and atomic forces[40]. About 600 surface configurations alone from equilibrated AIMD simulations at various surface temperatures ($T_s$=100~600 K) were also added into the initial training set. These points were fit to a preliminary PES, which enabled much more efficient quasi-classical trajectory calculations (see below) for highly vibrationally excited NO scattering from the surface. New data points from these trajectories were selectively included into the data set based on the same GED criterion so that they were not too close to existing points. The augmented data set was then used to update the PES, followed by a new iteration of trajectory calculations. This process was repeated until the state-to-state scattering probabilities was converged. To examine the influence of dispersion interaction between NO and LiF on the energy transfer dynamics, we recalculated energies and forces using the vdW-corrected optPBE-vdw functional[48] at those single point configurations, which were used to generate an independent neural network potential for NO/LiF(001). This functional resulted in a deeper adsorption well (-0.19 eV), but more or less the same potential energy landscape, yielding similar vibrational elasticity as the PW91-based one. (see Fig. S6).

DFT calculations for the NO/Au(111) system have been detailed elsewhere[35]. Briefly, the Au(111) surface was modeled by a four-layer slab in a 3 × 3 unit cell with the vacuum of 15 Å. The top two layers of Au atoms were allowed to move. The DFT settings were similar as above except that the Brillouin zone was sampled by a 4 × 4 × 1 Gamma $k$-points grid. Similar data sampling procedure was used and 3309 data points, consisting of 2722 points from previous work[35] and additional 587 surface configurations selected from equilibrated AIMD simulations of the bare Au(111) surface at various surface temperatures ($T_s$=100~1000 K).

**1. 2 Embedded Atom Neural Network**

Both PESs for the NO/LiF(001) and NO/Au(111) systems in this work have been constructed by means of our recently proposed embedded atom neural network (EANN)[41] approach. In the EANN representation, the total energy of the system is regarded as the sum of atomic energies, each of which is an output of an atomic neural network determined by the electron density of this atom embedded in the environment consisting of other atoms nearby[41],

$$E = \sum_{i=1}^{N} E_i = \sum_{i=1}^{N} NN_i\left(\boldsymbol{\rho}^i\right) \quad (1)$$

For simplicity, the embedded electron density like structural descriptors ($\boldsymbol{\rho}^i$) can be estimated from Gaussian-type orbitals (GTOs)[49] centered at neighboring atoms, resulting in multiple orbital-dependent density components,

$$\rho^i_{L,\alpha,r_s} = \sum_{l_x,l_y,l_z}^{l_x+l_y+l_z=L} \frac{L!}{l_x!l_y!l_z!} \left( \sum_{j=1}^{n_{atom}} c_j \varphi^{\alpha,r_s}_{l_x l_y l_z}(\mathbf{r}_{ij}) f_c(r_{ij}) \right)^2, \quad (2)$$

where $n_{atom}$ is the total number of atoms lying nearby the embedded atom within a cutoff radius ($r_c$) and $f_c(r_{ij})$ a cutoff function[50] to ensure that the contribution of each neighbor atom decays smoothly to zero at $r_c$. The GTO is written as,

$$\varphi^{\alpha,r_s}_{l_x l_y l_z}(\mathbf{r}_{ij}) = x^{l_x} y^{l_y} z^{l_z} \exp\left(-\alpha \left| r_{ij} - r_s \right|^2 \right), \quad (3)$$

where $\mathbf{r}_{ij}=(x, y, z)$ represents the Cartesian coordinates of the embedded atom $i$ with atom $j$ being the origin, $r_{ij}$ is the distance of atom $i$ and $j$, $l_x$, $l_y$ and $l_z$ represent the angular momentum components in each axis, and their sum is the total orbital angular momentum ($L$), $\alpha$ and $r_s$ are parameters that determine radial distributions of GTOs. Note that $c_j$ in Eq. (2) serves like an element-dependent expansion coefficient of an atomic orbital for atom $j$, which is optimized together with the element-dependent NN parameters. The EANN PES is invariant with respect to translation, rotation, and permutation[41]. The key advantage of this EANN method is that the density-like descriptors given in Eq. (2) scale linearly with respect to the number of neighboring atoms. It has been successfully applied to learn the PESs of molecules and materials[41,51], gas-surface reactions[42], as well as electronic friction tensors of adsorbates on surfaces[43].

The final NO/LiF(001) PES was fit to 3608 points with both energies and forces. These data points were divided into training and test sets with the ratio of 90:10. The hyperparameters of GTOs were $L$=0, 1, 2, $r_c$=6.0 Å, $\alpha$=0.6 Å$^{-2}$, and $\Delta r_s$=0.58 Å, resulting in 33 structural descriptors. Each atomic NN consists of two hidden layers with 50 and 60 neurons in each. The overall root-mean-square errors (RMSEs) are 19.4 meV for energies (0.29 meV/atom) and 16.6 meV/Å for atomic forces, respectively.

As for the NO/Au(111), 3309 data points with both energies and forces was fitted. The EANN architecture is the same as that for NO/LiF(001) except that the hyperparameters of GTOs were $L$= 0, 1, 2; $r_c$= 5.5 Å; $\alpha$= 0.8 Å$^{-2}$, $\Delta r_s$=0.5 Å, resulting in 36 structural descriptors. The overall RMSEs are 27.4 meV for energies (0.72 meV/atom) and 31.8 meV/Å for atomic forces, respectively.

Fig. S1 shows that stationary points for the NO adsorption on LiF(001) and Au(111) and dissociation on Au(111) only, have been well reproduced by the two EANN PESs. Fig. S2-S4 displays the potential energy landscapes in the two systems with respect to different coordinates. Fig. S5 indicates that the EANN PESs agree well with the DFT predicted surface phonon density of states.

### 1.3 Quasi-classical Trajectory Calculations

State-to-state quasi-classical trajectory (QCT) calculations were performed using the VENUS code heavily modified by us[52]. The trajectories were initiated at 6.0 Å above the LiF(001) and Au(111) surfaces with the molecular center randomly chosen to cover the $p(2\times2)$ LiF(001) unit cell and $p(3\times3)$ Au(111) unit cell. The diatomic molecule is treated as a rotating oscillator, whose internal energy was calculated semi-classically as a function of vibrational and rotational quantum numbers $v$ and $j$[53]. To compare with the experimental results[14, 17-19, 21], the NO molecule with different initial vibrational ($v_i$) and rotational states ($j_i$) and translational incidence energy $E_i$ along the surface normal was prepared mimicking experimental conditions. The propagation time step was 0.10 fs with velocity Verlet algorithm and up to 10 ps with the same termination criterion specified in the on-the-fly AIMD simulations described above. For scattered molecules, the vibrational action number $v_f$ was determined by Einstein−Brillouin−Keller (EBK) semi-classical quantization,[54] and rotational quantum number $j_f$ by the quantum mechanical expression for rotational angular momentum $\vec{J} = \sqrt{j_f(j_f+1)}\hbar$. In principle, NO is an open shell molecule whose rotational quantum number is supposed to be fractional due to the coupling between rotational and electronic angular momenta, which can however not be modeled by the current adiabatic dynamics calculations. As a result, the initial rotational quantum number was set as the nearest integer of the experimental value, which should have negligible influence on the results. To obtain the final ro-vibrational state distributions, the fractional vibrational and rotational quantum numbers were binned into the nearest integers. This histogram binning has been found to yield quite similar results as the Gaussian binning in the NO/Au(111) system[36]. Over 10000 trajectories were computed at each incidence energy to converge the final dynamics results with good statistics.

Table S1. Average rotational, vibrational, translational, and total energy losses (in eV) of NO before and after scattering from LiF(001) and Au(111) with top two layers of surface atoms relaxed and fixed. Initial conditions include NO($v_i$=16) at $E_i$=0.52 eV and $T_s$=300 K for Au(111), and NO($v_i$=12) at $E_i$=0.42 eV and $T_s$=290 K for LiF(001), respectively.

| Mean energy loss (eV) | Au(111) | | LiF(001) | |
|---|---|---|---|---|
| | relaxed | fixed | relaxed | fixed |
| $<\Delta E_{rot}>$ | 0.40 | 0.49 | 0.035 | 0.066 |
| $<\Delta E_{vib}>$ | -1.0 | -0.71 | ~0 | ~0 |
| $<\Delta E_{trans}>$ | -0.036 | 0.22 | -0.24 | -0.067 |
| $<\Delta E_{tot}>$ | -0.65 | 0 | -0.20 | 0 |

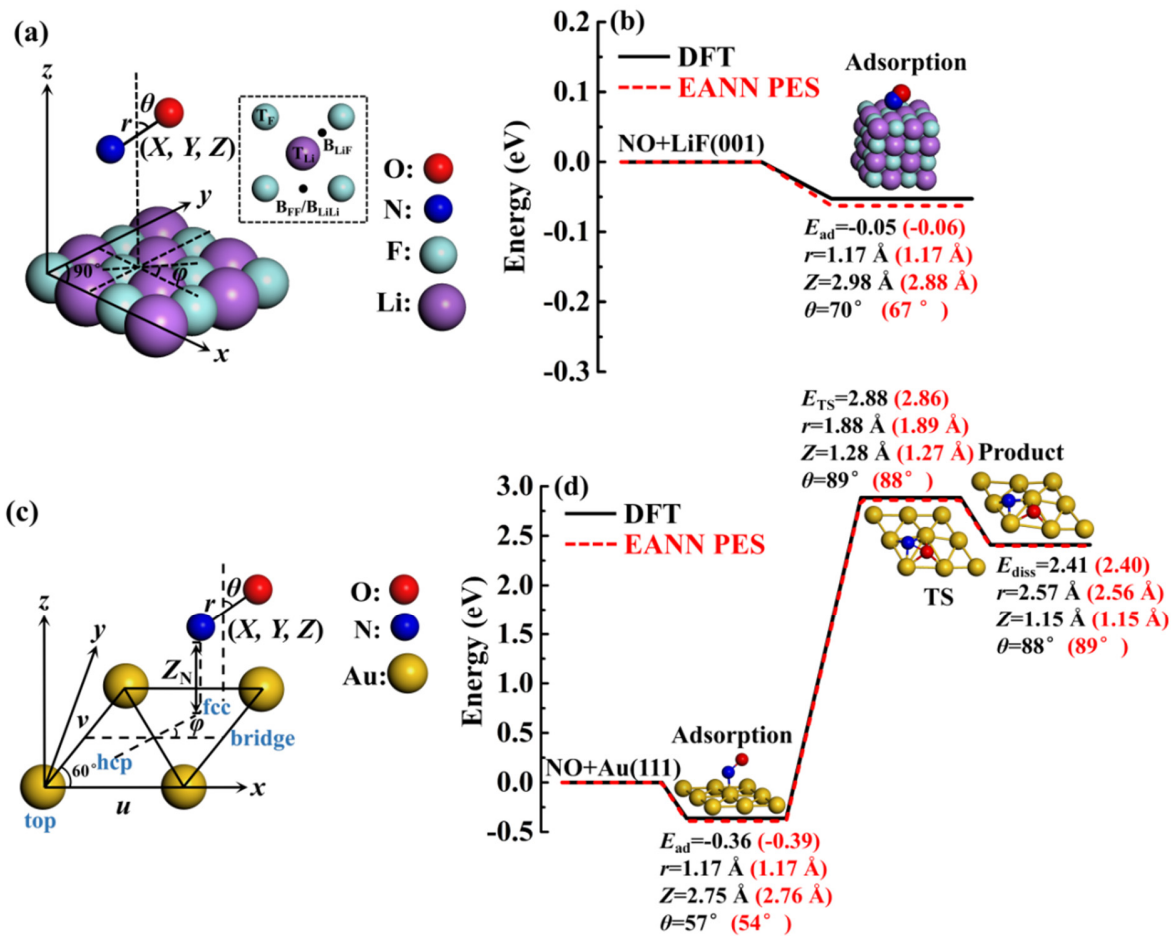

Figure S1. (a)(c) Jacobi coordinates for describing the NO/LiF(001) and NO/Au(111) system. The inset in (a) shows high-symmetry sites on the LiF(001) surface. (b) Comparison of geometries and energies (in eV) of stationary points in the NO/LiF(001) system, obtained from DFT (black) and EANN PES (red). (d) The same as (b) but in the NO/Au(111) system, where the dissociation transition state (TS) and the dissociated product are included.

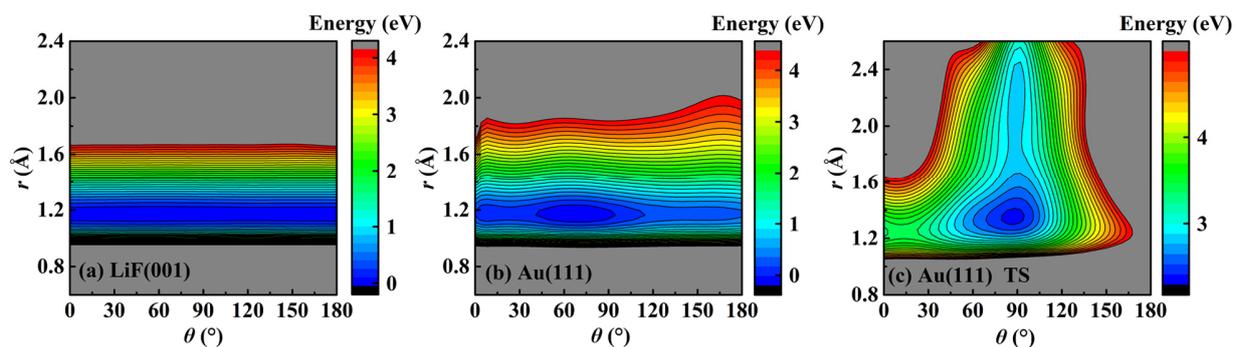

Figure S2. (a)(b) Two-dimensional contours of the NO/LiF(001) and NO/Au(111) EANN PESs as a function of $r$ and $\theta$, with other coordinates of the molecule (defined in Fig. S1) and surface atoms fixed at the respective adsorption states, respectively. (c): same as (b), except that other coordinates of the molecule and surface atoms are fixed at the dissociation transition state in the NO/Au(111) system.

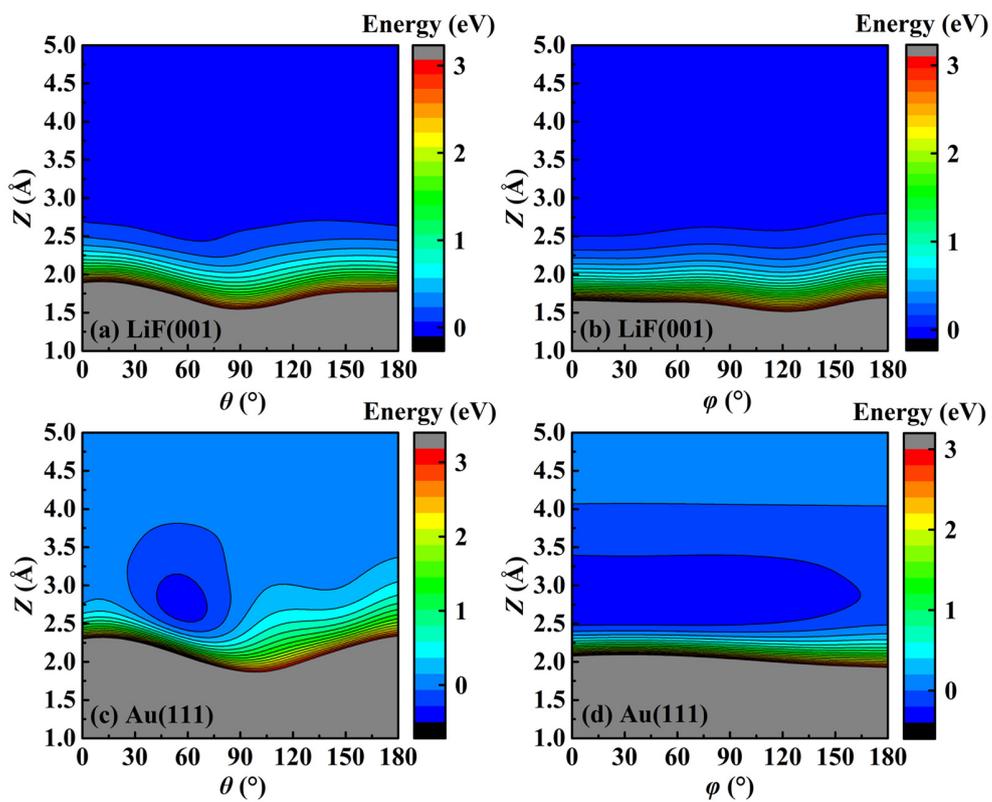

Figure S3. Two-dimensional contours of the EANN PESs as a function of $Z$ and $\theta$ (a-b), or as a function of $Z$ and $\varphi$ (c-d), with other coordinates of the molecule and surface atoms fixed at the respective adsorption states.

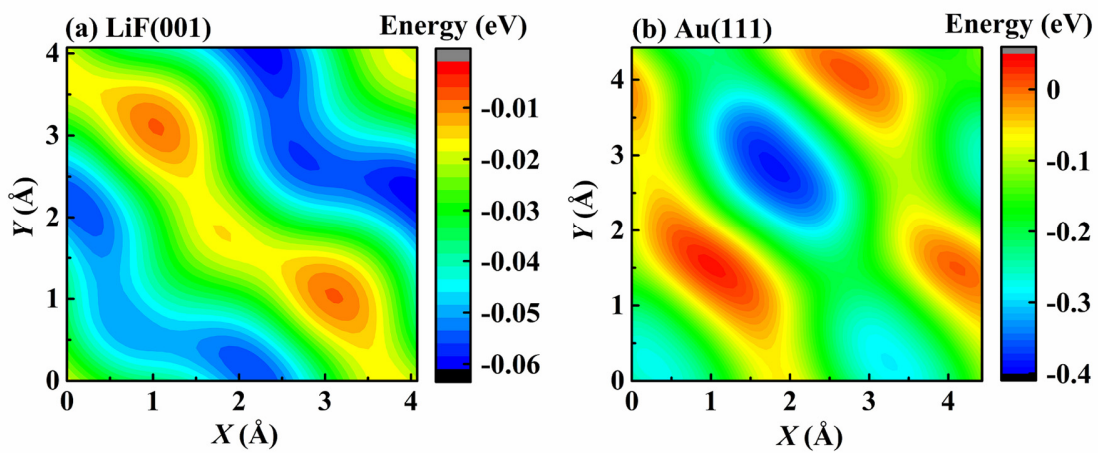

Figure S4. Potential energy landscapes of the EANN PESs as a function of *X* and *Y*, with other coordinates of the molecule and surface atoms fixed at the respective adsorption states.

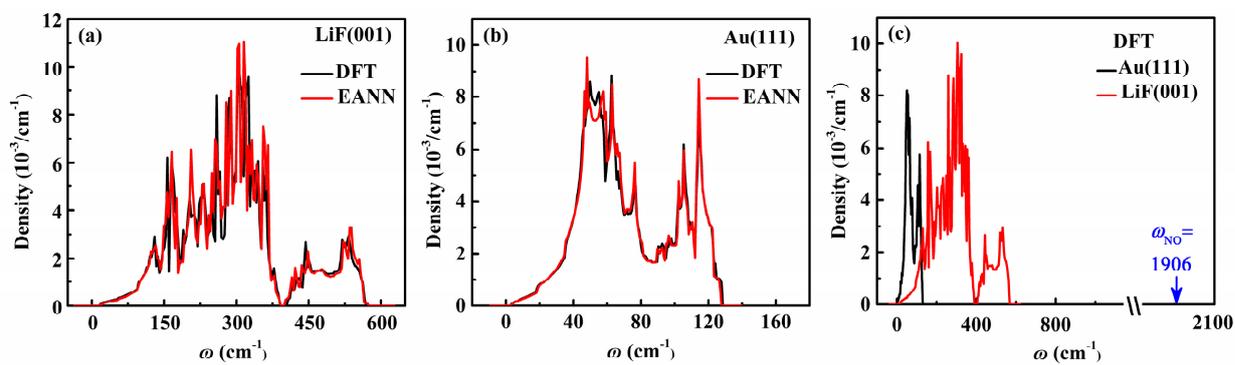

Figure S5. (a)(b) Comparison of phonon density spectra of the LiF(001) and Au(111) surfaces obtained by DFT(black) and EANN PES (red). (c) Comparison of DFT phonon frequencies LiF(001) (red) and Au(111) (black), with the harmonic frequency of the NO molecule (blue arrow).

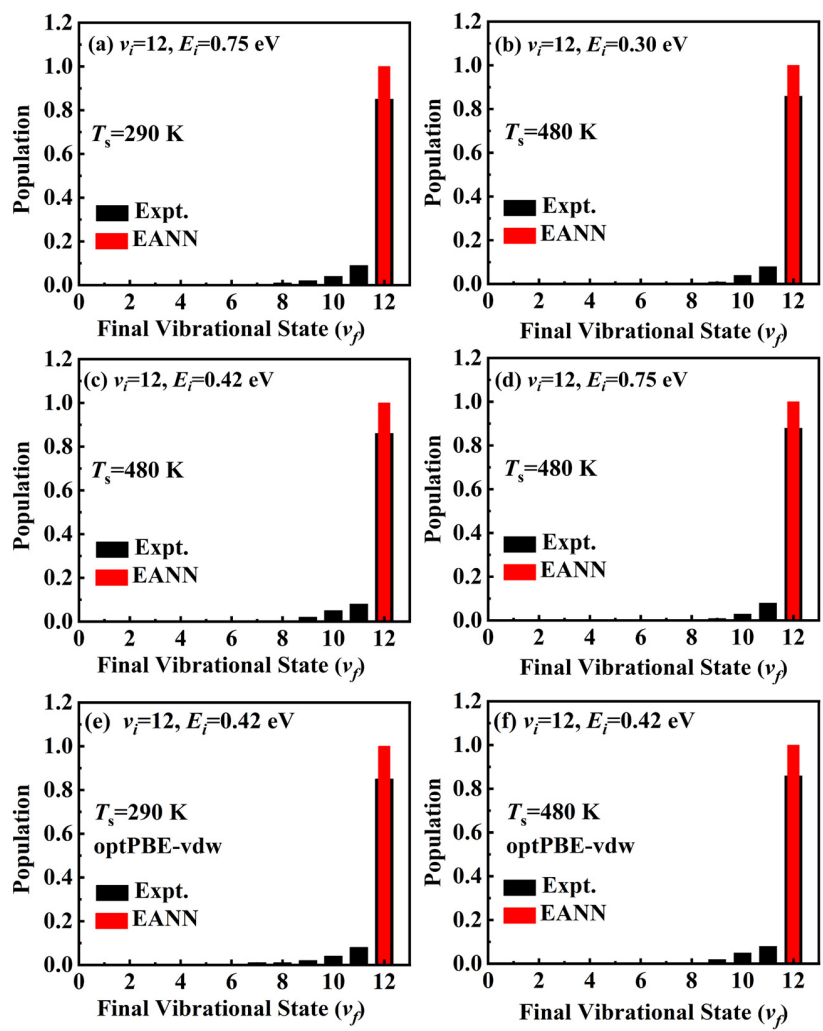

Figure S6. Comparison of experimental (black) and calculated (red) vibrational state distributions of highly vibrationally excited NO scattered from LiF(001). Initial conditions are selected according to experiment, including (a) NO($v_i$=12, $E_i$=0.30 eV) with $T_s$=290 K; (b-d) NO($v_i$=12, $E_i$=0.30 eV, $E_i$=0.42 eV and $E_i$=0.75 eV) with $T_s$=480 K. Also compared in panels (e-f) are calculated results on the optPBE-vdw density functional based EANN PES and the experimental data for NO($v_i$=12, $E_i$=0.42 eV) with $T_s$=290 K and $T_s$=480 K, respectively. This PES has a deeper adsorption well (-0.19 eV) than the PW91-based one (-0.06 eV), which however does not alter the vibrational relaxation dynamics.

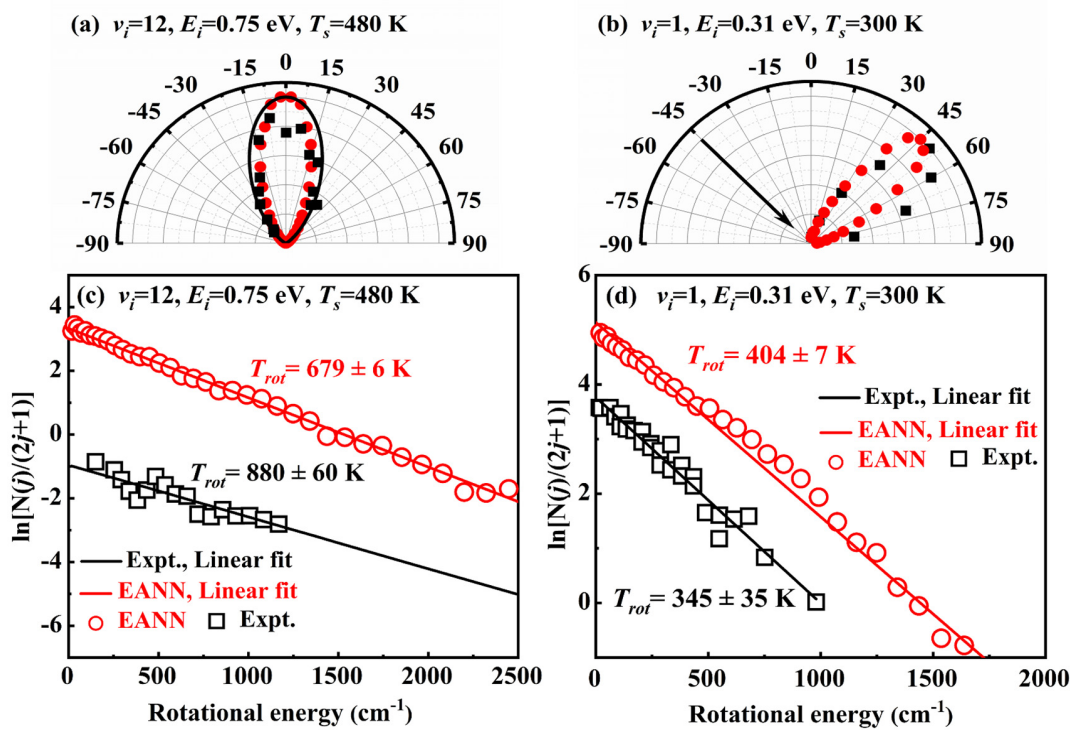

Figure S7. (a-b) Comparison of experimental (black squares) and theoretical (red circles) angular distributions for NO vibrationally elastic scattering from LiF(001). The $\cos^6(\theta)$ function (black line) in (a) is the experimental fit to guide eyes. (c-d) Comparison of experimental (black squares) and calculated (red circles) Boltzmann factors of final rotational states of vibrationally elastic scattering of NO from LiF(001). Linear fits of experimental (black line) and calculated (red line) results yield the corresponding rotational temperatures ($T_{rot}$). Two experimental conditions are considered, namely (a)(c) NO($v_i$=12) with normal incidence at $E_i$=0.75 eV and $T_s$=480 K; (b)(d) NO($v_i$=1) with an incidence angle of 45° at $E_i$=0.31 eV and $T_s$=300 K.

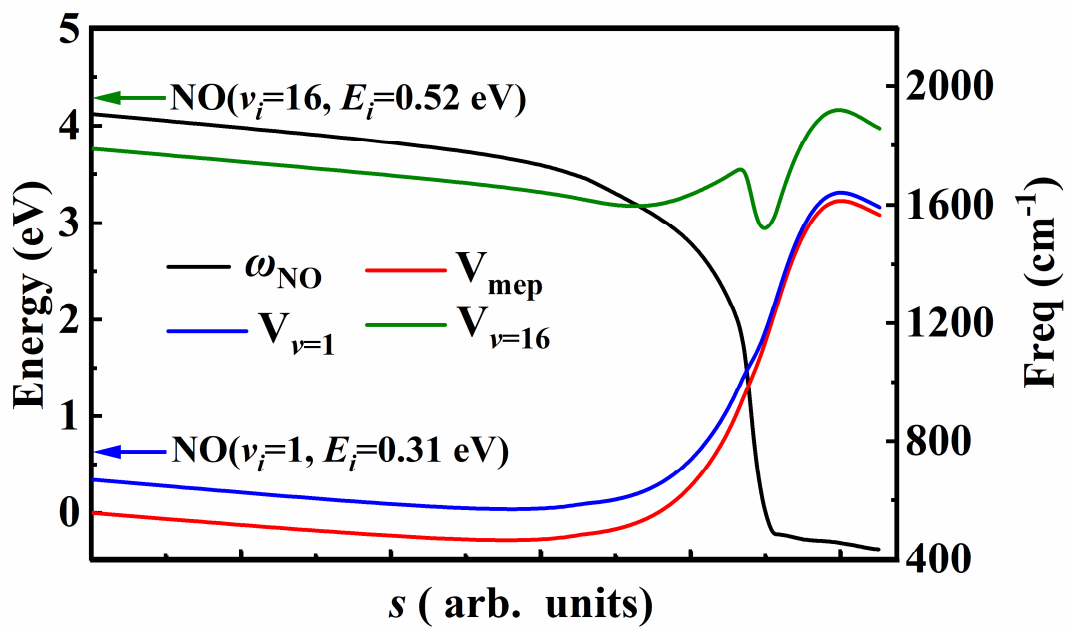

Figure S8. Adiabatic evolution of the NO vibrational frequency (black) and potential energy ($V_{mep}$, red) along the minimum energy path of NO dissociation on Au(111). Corresponding vibrationally adiabatic potentials for NO($v_i$=1) (blue) and NO($v_i$=16) (green) dissociation.

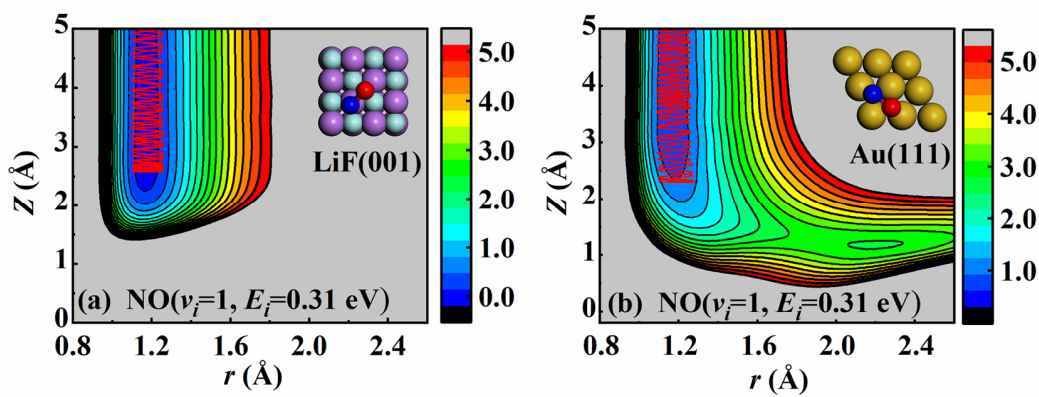

Figure S9. Two-dimensional cuts of (a) the NO/LiF(001) and (b) the NO/Au(111) PESs as a function of the N-O distance ($r$) and the molecular height ($Z$) above the surface, with other coordinates fixed at the adsorption state on LiF(001) and the dissociation transition state on Au(111). A representative trajectory (red line) of each system is projected on top of the corresponding PES, for NO($v_i$ =1) at $E_i$=0.31 eV on Au(111) and LiF(001) surfaces with $T_s$=300K, the incidence angle was 45°.

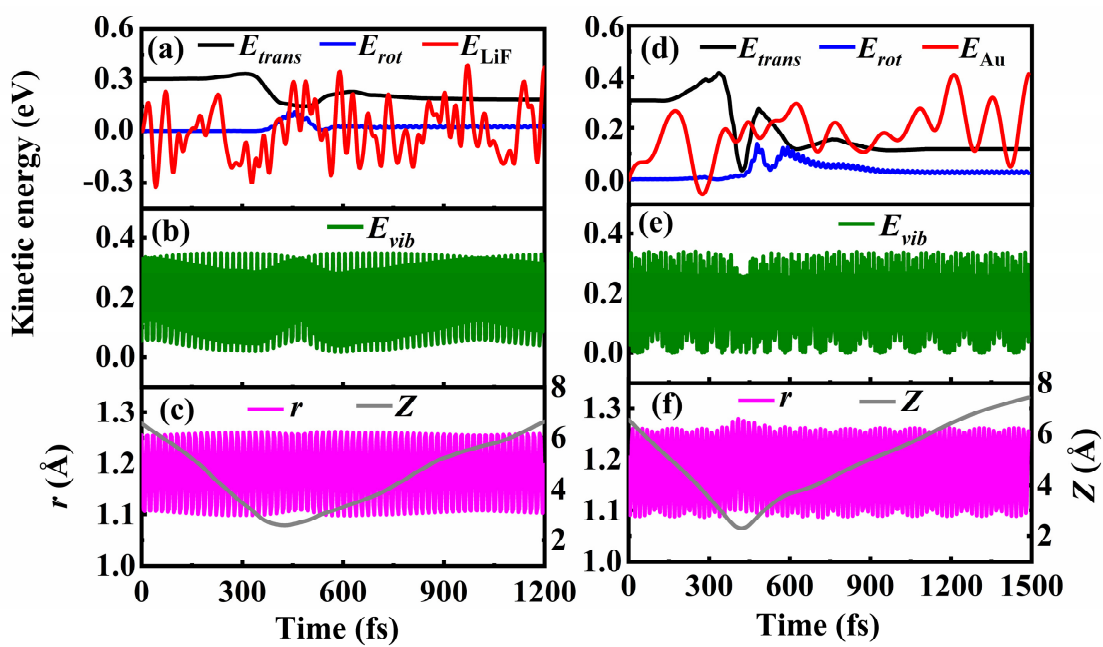

Figure S10. Kinetic energy and geometric evolutions as a function of time during a representative scattering trajectory for NO($v_i$=1, $E_i$=0.31 eV) from (a-c) LiF(001) (d-f) Au(111), including the kinetic energy in NO translation ($E_{trans}$, black), rotation ($E_{rot}$, blue), and vibration ($E_{vib}$, green), the average kinetic energy of surface atoms relative to the initial value ($E_{LiF}$ and $E_{Au}$, red), the N-O distance ($r$, magenta), and the molecular height ($Z$, gray).

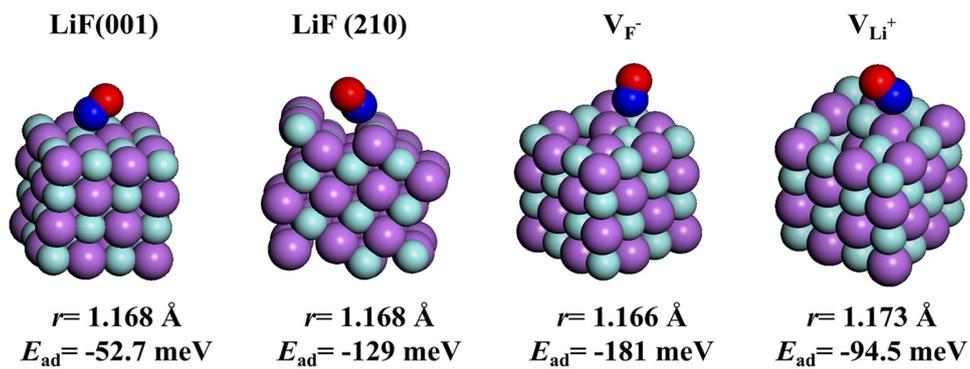

Figure S11. Comparison of binding energies of NO and the adsorbed NO bond lengths on the pristine and several representative defective LiF(001) surfaces, e.g. LiF(210), LiF(001) with a single $F^-$ vacancy, or a single $Li^+$ vacancy. The color scheme here is the same as Figure S1.

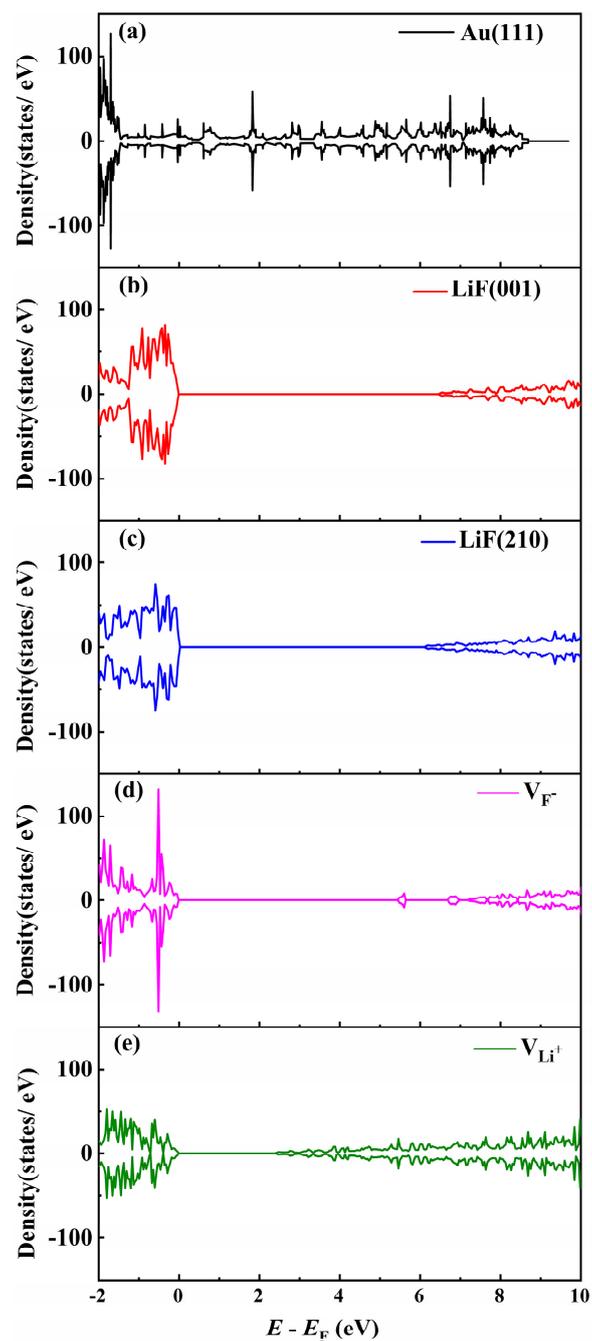

Figure S12. Comparison of density of electronic states of Au(111) (black), LiF(001) (red), LiF(210) (blue), and LiF(001) with a single $F^-$ vacancy (magenta), and a single $Li^+$ vacancy (green).

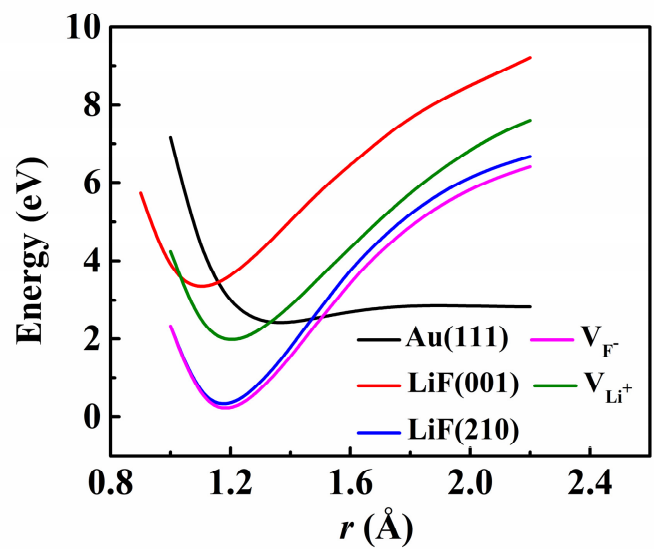

Figure S13. Adiabatic potential energy curves of NO on Au(111) (black), LiF(001) (red), LiF(210) (blue), and LiF(001) with a single F$^-$ vacancy (magenta) and Li$^+$ vacancy (green) as a function of $r$ to show the possibility of NO dissociation on these surfaces. For this purpose, the molecular orientation is kept as that at the transition state on Au(111) and molecular height $Z$ is fixed at 1.28 Å for NO on Au(111) and 1.50 Å for NO on LiF(001), LiF(210), and LiF(001) including F$^-$ vacancy and Li$^+$ vacancy.